\documentclass[aps,prb,twocolumn,showpacs,preprintnumbers,amsmath,amssymb,superscriptaddress]{revtex4}

\usepackage{graphicx}
\usepackage{dcolumn}
\usepackage{bm}
\newcommand{\nc}{\newcommand}
\nc{\be}{\begin{equation}}
\nc{\ee}{\end{equation}}
\nc{\bea}{\begin{eqnarray}}
\nc{\eea}{\end{eqnarray}}
\nc{\bean}{\begin{eqnarray*}}
\nc{\eean}{\end{eqnarray*}}
\nc{\mb}{\mbox}
\nc{\rnc}{\renewcommand}
\nc{\BPi}{\mb{\boldmath$\Pi$}}
\nc{\Bp}{\mb{\bf p}}
\nc{\BA}{\mb{\bf A}}
\nc{\vk}{\mb{\bf k}}
\nc{\vp}{\mb{\bf p}}
\nc{\vn}{\mb{\bf n}}
\nc{\vq}{\mb{\bf q}}
\nc{\rr}{\mb{\bf r}}
\nc{\vz}{\hat {\mb{\bf z}}}
\nc{\vj}{\mb{\boldmath$j$}}
\nc{\vg}{\mb{\boldmath$g$}}
\nc{\x}{\mb{\boldmath$x$}}
\nc{\A}{\mb{\boldmath$A$}}
\nc{\va}{\mb{\boldmath$a$}}
\nc{\vs}{\mb{\boldmath$\sigma$}}
\nc{\vpi}{\mb{\boldmath$\pi$}}
\nc{\nab}{\nabla}
\nc{\X}{\sf x}

\begin{document}

\title{
Aharonov-Casher and spin Hall effects in two-dimensional mesoscopic ring structures with strong spin-orbit interaction
}
\author{M. F. Borunda}
\affiliation{Department of Physics, Texas A\&M University, College Station, TX
77843-4242, USA}
\author{Xin Liu}
\affiliation{Department of Physics, Texas A\&M University, College Station, TX
77843-4242, USA}
\author{Alexey A. Kovalev}
\affiliation{Department of Physics, Texas A\&M University, College Station, TX
77843-4242, USA}
\author{Xiong-Jun Liu}
\affiliation{Department of Physics, Texas A\&M University, College Station, TX
77843-4242, USA}
\author{T. Jungwirth}
\affiliation{Institute of Physics  ASCR, Cukrovarnick\'a 10, 162 53 Praha 6, Czech Republic }
\affiliation{School of Physics and Astronomy, University of Nottingham, Nottingham NG7 2RD, UK}
\author{Jairo Sinova}
\affiliation{Department of Physics, Texas A\&M University, College Station, TX
77843-4242, USA}
\affiliation{Institute of Physics  ASCR, Cukrovarnick\'a 10, 162 53 Praha 6, Czech Republic }
\date{\today}

\begin{abstract}
We study the quantum interference effects induced by the Aharonov-Casher phase in asymmetrically confined two-dimensional electron and heavy-hole ring structures systems taking into account the electrically tunable spin-orbit (SO) interaction. We have calculated the non-adiabatic transport properties of charges (heavy-holes and electrons) in two-probe thin ring structures and compare how the form of the SO coupling of the carries affects it. 
We show that both the SO splitting of the bands and the carrier density can be used to modulate the conductance through the ring.
We show that the dependence on carrier density is due to the backscattering from the leads which shows pronounce resonances when the Fermi energy is close to the eigenenergy of the ring.
We also calculate the spin Hall conductivity and longitudinal conductivity in four-probe rings as a function of the carrier density and SO interaction, demonstrating that for heavy-hole carriers both conductivities are larger than for electrons. Finally, we investigate the transport properties of mesoscopic rings with spatially inhomogeneous SO coupling. We show that devices with inhomogeneous SO interaction exhibit an electrically controlled spin-flipping mechanism.
\end{abstract}
\maketitle

\noindent
\section{Introduction}
In recent decades, advances in technology have allowed the fabrication of electronic components of mesoscopic dimensions. Given the nanometer spatial confinement of the charge carriers, some of the important features exhibited are due entirely to quantum mechanical effects. Notwithstanding the future applications that will exploit such effects, several areas of fundamental physics will benefit from the study of such devices.\cite{Imry_book} 

One such area is the study of geometric phases.\cite{Anandan} When a quantum particle undergoes cyclic evolution motion in the system's parameter space, it acquires a geometric phase that will strongly influence the transport properties of the system. Pancharatnam,\cite{Pancharatnam} when studying polarized light in crystals, found from interference experiments that the path that the light travels (or the sequence of measurements performed during that path) was responsible for an additional phase component. In the same manner, the path taken by a beam of electric charges is important when in the presence of electromagnetic vector potentials. Even in the case of the potential not producing a field that results in a force acting on the particles, the particles may gain a quantum phase that depends on the path traversed. Examples of such phase gains are the Aharonov-Bohm (AB) effect and its relativistic cousin the Aharonov-Casher (AC) effect. In the AB effect the particle gains a phase as it moves in a path enclosing a magnetic flux.\cite{Aharonov} In the AC effect the phase acquired follows readily from spin-orbit (SO) coupling instead of a magnetic field.\cite{PRL_53_319} The generalized explanation to these phase dependent phenomena in adiabatically and non-adiabatically evolving quantum systems was given by Berry\cite{Berry} and by Aharonov and Anandan,\cite{PRL_58_1593} respectively. 

These geometric phases can be studied most readily in mesoscopic ring structures. Studies on mesoscopic rings with inhomogeneous magnetic fields showed analogies between geometric phases due to SO interaction and the phases acquired by moving electrons in an effective inhomogeneous magnetic field with opposite sign for each spin.\cite{Loss_1990} At the same time, it was found that SO interaction shifts the AB oscillations and adds destructive interference to the conducting rings.\cite{Aronov_1993}

Applications arising from tuning the phase of the carriers have been widely discussed,\cite{APL_75_695,PRB_69_235310,PRB_69_155335, Souma, Souma:2004_a,PRB_76_155326} starting with the proposal by Nitta {\it et. al.} of an spin-interference device to modulate the current flow.\cite{APL_75_695} 
The  proposed device consists of a paramagnetic electrode lead injecting charge into a ring section fitted with a gate and another lead that would extract the charge carriers. In narrow-gap semiconductor systems (electron or heavy hole)  the SO coupling strength can be controlled by applying a gate voltage.\cite{SOC_control} Thus, the phase difference acquired in each of the arms of the ring and the ensuing interference effect can be modified by the gate voltage which results in periodic oscillations of the conductance.\cite{APL_75_695}

Recent experiments have confirmed that a gate bias modifies the oscillations in the magneto-conductance curves which demonstrates gate-controlled changes in the geometric phase.\cite{Konig,Koga_2006,Habib} The magneto-conductance oscillations as a function of both magnetic field and gate voltage that controls the SO interaction strength in HgTe ring structures have been studied by Konig and coworkers.\cite{Konig} Their measurements exhibit a nonmonotonic phase change as a function of the gate voltage and establish the connection between this observation and the AC effect by finding a quantitative agreement between their experimental results and Landauer-B{\"u}ttiker numerical calculations of a multichannel ring with Rashba SO interaction.\cite{Konig} Koga {\it et. al.} measured the sheet conductivity of square loops arrays observing a gate voltage mechanism influencing the spin interference of electrons.\cite{Koga_2006} Habib {\it et. al.} also measured resistance oscillations in a two-dimensional hole ring structure.\cite{Habib} The oscillations depend on a front gate voltage but are not completely attributed to SO splitting due to asymmetry in the structure and the low density of the system.\cite{Habib} Similarly AB oscillations in the magneto-conductance have been measured in p-type ring structures lacking a gate to control the SO splitting but in systems with considerable SO splitting ($\Delta_{SO}/E_F \sim 0.3$) where the SO induced field is reported to be as strong as $B_{eff} = 0.25 T$.\cite{PRL_99_176803}

In the theoretical front the modulation of the electric current driven by quantum interference effects in 1D coherent electron systems was calculated analytically and compared to numerical simulations of a 2D ring in which only one mode is conducting.\cite{PRB_69_235310,PRB_69_155335} Both calculations confirmed that the spin-dependent transport results in strong quasiperiodic modulations to the conductivity. 
Moln{\'a}r and coworkers calculated the conductance as a function of the Fermi wave vector of the incident electrons and the SO coupling in the mesoscopic rings.\cite{PRB_69_155335} Souma and Nikoli{\'c} numerically compared the conductance modulation of the mesoscopic rings as more channels open for conduction and they found that the modulation pattern remains but is affected as more modes become available. Moreover, they concluded that the spin-interference of the channels is not cumulative, given that for single-channel devices the conductivity can be null at certain values of the SO interaction and the same does not hold in the multi-channel devices.\cite{Souma} In addition, Souma and Nikoli{\'c} calculated the spin-Hall conductance of ring structures and found that the spin Hall conductance is also modulated by the SO interaction.\cite{Souma:2004_a} In their proposed four-probe ring, a pure spin current is induced in the transverse probes when unpolarized current flows in the longitudinal probes.  A recent calculation also considers the effect of an inhomogeneous SO coupling in the 2DEG ring structures.\cite{PRB_76_155326} By mapping the SO interaction to a spin dependent magnetic field, Tserkovnyak and Brataas found that in the weak SO regime quantum-interference effects can be stronger due to the inhomogeneity of the field.\cite{PRB_76_155326}

In this paper, we revisit the problem of transport characteristics of ring structures in electron and hole doped system. In section \ref{Ham} we present  Hamiltonians that describe electron and heavy-hole systems in thin ring geometries and outline the two methods used to calculate their transport properties. In section \ref{hh_vs_e} we confirm numerically analytical results\cite{Kovalev:2007_a} for a single channel ring embedded in a narrow quantum well. The first part of the section focuses on the consequences for the conductance of the change in the Hamiltonian from wave-vector $k$-linear spin splitting term compared to a $k$-cubic term. In the second part of the section we explain the results and the dependence on the carrier concentration. In the last part of section \ref{hh_vs_e} we present the calculation of the spin Hall conductivity in a four probe ring geometry for heavy-hole carriers and compare it to the conductivity obtained for electrons.  Section \ref{inhomo} explores numerically the effect of inhomogeneous SO in mesoscopic rings. We study how the AC effect is affected by having a region with no SO coupling within the ring and show increases in the strength of the signals obtained from the conductivities. In this section we also demonstrate that devices with inhomogeneous SO interaction exhibit an electrically controlled spin-flipping mechanism not present in the case of homogeneous SO coupling. In section \ref{conc} we summarize the results and present our conclusions.

\section{Model Hamiltonians and Calculation Methods}\label{Ham}
\subsection{Hamiltonians}
\begin{figure}[!t]
\begin{center}
\includegraphics[width=0.5\textwidth]{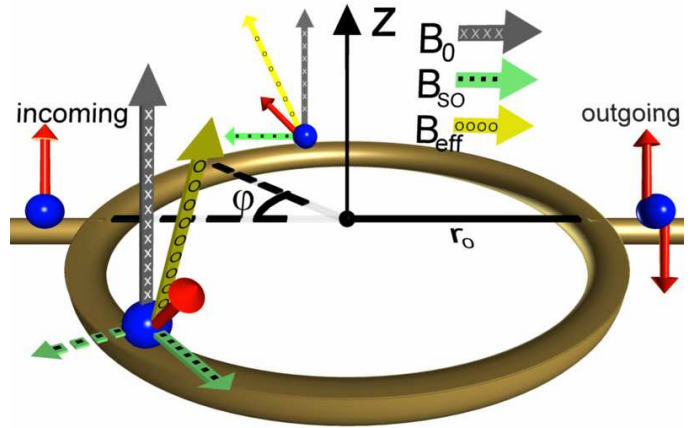}
\caption{
(Color online) Semiconductor quantum well patterned as a ring of radius $r_0$ in the presence of a magnetic field $B_0$ and of spin-orbit coupling. An electron(hole) spin traveling around the ring acquires phase due to the sum of the magnetic fields acting on it. The effective field, $B_{eff}$, is given by the yellow(o markings) arrow, the applied out-of-plane magnetic field, $B_0$, is represented by the gray(x markings) arrow, and the momentum dependent in-plane magnetic field due to spin-orbit interaction, $B_{SO}$, seen in green($\bullet$ markings) arrow. The orientation of the spin-orbit field changes at different rates depending on the carriers, due to the holes (solid arrow) having a cubic in momentum spin splitting and the electrons (dashed arrow) having a spin splitting that depends linearly on the momentum.
}
\label{f:ring1}
\end{center}
\end{figure}
In two-dimensional systems the effective mass Hamiltonian in the presence of both SO coupling and a perpendicular magnetic field, $B_z$, is:
\be
\label{H}
H = \frac{\BPi^2}{2 m} + H_z + H_{conf}+ H_{SO}  
\ee
where $\BPi = \Bp + (e/c) \BA $, $H_z = \frac{1}{2} g \mu_g \sigma_z B_z$, and the electrostatic confining potential is given by $H_{conf}$. The SO terms in the single particle Hamiltonian is given by $H_{SO}^e  =  \alpha_1 \left( \hat{\sigma} \times \BPi \right)_z / \hbar$ for electrons and $H_{SO}^{hh} = \alpha_3 \left( \hat{\sigma}_+ \Pi^3_- - \sigma_{-} \Pi^3_+ \right)/\hbar^3$ for heavy holes. We limit our study to Rashba type SO coupling in narrow quantum wells, i.e. the structure inversion asymmetry is responsible for SO. The strength of the spin-orbit interaction is given by $\alpha_n$. The system is illustrated in Fig. \ref{f:ring1}.

As outlined in Ref. \onlinecite{PRB_66_033107}, to obtain the effective 1D Hamiltonian and due to the subtleties introduced by the SO interactions, it does not suffice to discard the derivatives in the radial direction and set $r=r_0$ in the 2D Hamiltonian. The single particle 1D Hamiltonian is found by assuming that the confining potential of the 2D system is such that the electron wave functions are confined to a 1D ring:\cite{PRB_66_033107}
\be
\label{H_1}
H^e(\varphi,\Phi) = \frac{(\hbar ~ \tilde{\partial}_\varphi)^2}{2m^* r_0^2}  + H_z 
- \frac{\hbar^2 Q_e}{2 m^* r_0^2} \left(\hat{\sigma}_{1,r} \tilde{\partial}_\varphi - \frac{i }{2 } \hat{\sigma}_{1,\varphi}\right) 
\ee
where $Q_e\equiv 2m^* \alpha_1 r_0/\hbar^2$ characterized the strength of the spin-orbit coupling splitting in the thin ring limit relative to the angular leading term.
The same process was used in finding the effective 1D heavy-hole Hamiltonian:\cite{Kovalev:2007_a}
\bea
\label{H_3}
H^{hh}(\varphi,\Phi) &=& \frac{(\hbar ~ \tilde{\partial}_\varphi)^2}{2m^* r_0^2} 
+ \frac{\alpha_3}{r_0^3} \hat{\sigma}_{3,r} \left( (\tilde{\partial}_\varphi)^3 + \tilde{\partial}_\varphi \left[\frac{3 r_0^3}{w^2} - \frac{7}{2} \right]\right) \nonumber \\ &&  
+ \frac{i \alpha_3}{2 r_0^3} \hat{\sigma}_{3,\varphi} \left( \frac{3 r_0^3}{w^2} + 1 + 3 (\tilde{\partial}_\varphi)^2 \right) + H_z
\eea
with $r_0$ being the radius of the ring, $w$ is the half width of the ring channel, $m^*$ the effective particle mass, $\varphi$ the angular coordinate, $\tilde{\partial}_\varphi= i \frac{\partial}{\partial \varphi} + \Phi$, the Pauli spin operators ($\hat{\sigma}_x$ and $\hat{\sigma}_y$) assume their usual values. To allow a cleaner notation, we use the following generalization of the Pauli spin operators in cylindrical coordinates: $\hat{\sigma}_{n,r} = \cos (n \varphi) ~ \hat{\sigma}_x + \sin (n \varphi) ~ \hat{\sigma}_y$ and $\hat{\sigma}_{n,\varphi} = \cos (n \varphi) ~ \hat{\sigma}_y - \sin (n \varphi) ~ \hat{\sigma}_x$. The magnetic field is included by its corresponding magnetic flux, $\Phi = \pi r_0^2 B/\Phi_0$, where $\Phi_0 = h c /e$ is the magnetic flux quantum. We model the heavy hole character of the band in that the SO interaction is cubic in momentum and the 2D Hamiltonian contains terms with $r^n$ and nth derivatives with respect to $r$, $n = 1,2,3$. 
 The projection of the lowest radial solution of the 2D Hamiltonian into a 1D Hamiltonian has as a consequence that in the heavy-hole calculation there are terms that depend on the width of the channel. As a contrast, in the electron system there are only terms dependent on the first power of $r$ and the first derivatives with respect to $r$, thus the desired 1D Hamiltonian does not have the dependence on the width. In order to have a fully 1D systems the radius of the ring has to be larger than the width of the arms. In the electron systems the limit $w =0$ can be taken; this is not possible in the heavy-hole systems. Nevertheless, we can obtain a simpler Hamiltonian than the one in Eq. (\ref{H_3}) by assuming that the width of the channel is of the order of the Fermi wavelength ($k_f w \lesssim 1$):\cite{Kovalev:2007_a}
\be
\label{H_3_simple}
H^{hh}(\varphi,\Phi) = \frac{(\hbar ~ \tilde{\partial}_\varphi)^2}{2m^* r_0^2} + H_z
+ \frac{\hbar^2 Q_{hh}}{2m^*r_0^2}\left( \hat{\sigma}_{3,r} \tilde{\partial}_\varphi + \frac{3i}{2} \hat{\sigma}_{3,\varphi} \right)  
\ee
where $Q_{hh}\equiv 6\alpha_3 m^* r_0/(\hbar^2 w^2)$ characterized the strength of the spin-orbit coupling splitting in the thin ring limit relative to the angular leading term and controls the precessions angle over the circumference of the ring.
If the lateral confinement is not very strong leading to thick rings (the width of the channel is comparable to the Fermi wavelength), then we are not truly in the lowest radial mode. Although from Eq. (\ref{H_3}) it is possible to write a Hamiltonian where the terms $w^{-2}$ are not dominant, this will not be realistic unless those higher radial modes are taken into account. Unfortunately, current experiments have not reached the limit of single channel rings. We also consider structures where the SO interaction varies along the azimuthal direction. Then an additional term appears in the Hamiltonian:
\bea
\label{H_1_vary}
H^e(\varphi,\Phi) &=& -\frac{\hbar}{2m^* r_0^2} {\tilde\partial}_\varphi^2 - \frac{i}{r_0} \alpha_1(\varphi) \left(\hat{\sigma}_{1,r} {\tilde\partial}_\varphi + \frac{1}{2}\hat{\sigma}_{1,\varphi} \right) \nonumber \\ && - \frac{i\hat{\sigma}_{1,r}}{2r_0} \frac{\partial \alpha_1(\varphi)}{\partial \varphi},
\eea
Our calculations do not consider the mixing of heavy-hole and light-hole states that is induced by the confinement. This neglected coupling has been shown to be responsible for an additional energy dependent non-adiabatic phase.\cite{prb_77_193304}   
\subsection{Landauer-B\"{u}ttiker formalism}
Numerical modeling of the rings is performed using the Landauer-B\"{u}ttiker formalism.\cite{Souma,Buttiker_prl_1986,datta} We assume that the ring is attached to semi-infinite paramagnetic leads that act as reservoirs for the quasiparticles. The procedure is as follows. First, the 1D Hamiltonians of the ring structure are discretized in a tight-binding model,\cite{Aldea_1992} and used to find the retarded/advanced Green's function:
\be
G^{R/A}(E) = (E - H - \Sigma^{R/A})^{-1}
\ee
The last term is the self-energy ($\Sigma^{R/A}$). It holds the connection between the structure and the semi-infinite leads. $\Sigma^R = \sum_{p} \Sigma_{p}^R$ involves a sum over all leads because we are assuming that all the leads are independent and thus their effects are additive. The description of the probes attached to our sample is made by solving the Green's function of a semi-infinite strip analytically. To do so, we need to calculate the self-energy terms which involves solving for the wave functions of the quasiparticles flowing from each lead into and out of the sample. 
With both the Green's functions and the self-energies at hand, we can write the transmission function:
\be
T_{pq} = \text{Tr}\left[\Gamma_p G^R \Gamma_q G^A \right]
\ee
where
\be
\Gamma_p = i \left[\Sigma_{p}^R - \Sigma_{p}^A\right]
\ee
The Green's function describes the dynamics of the charge carrier in the conductor taking the leads into account where $\Gamma_p$ represents the strength of the coupling of the leads to the sample. The total current flowing in each lead is obtained from the Landauer-B\"{u}ttiker formula:\cite{Buttiker_prl_1986,datta}
\be
I_p = \frac{e^2}{h} \sum_{q \neq p} T_{pq}(V_p - V_q)= I_p^{\uparrow} + I_p^{\downarrow}
\ee
The spin current can be defined as:\cite{Pareek_prl_2004,Nikolic:2004_a}
\be
I_{p,\sigma}^{\text{spin}} = \frac{\hbar}{2e}(I_p^{\uparrow} - I_p^{\downarrow}) = \frac{e}{4 \pi} \sum_{q \neq p, \sigma'} T_{pq}^{\sigma \sigma'}(V_p - V_q)
\ee
where $\sigma, \sigma'$ are the spin indices and the transmission function $T_{pq}^{\sigma \sigma'}$ gives us the probability of a particle with spin $\sigma'$ injected in lead $q$ is extracted from lead $p$ with spin $\sigma$. In the four probe structure current is injected in the right lead and extracted in the left lead. The transverse leads (top and bottom) act as voltage probes. Thus, the longitudinal and spin Hall conductance are:
\bea
G_L &=& \frac{I_R}{V_L - V_R} \label{e:G_l} \\
G_{sH} &=& \frac{I_{T,\uparrow}^{\text{spin}} - I_{T,\downarrow}^{\text{spin}}}{V_R}
\eea 
\subsection{Boundary condition tight-binding model}
\begin{figure}[ht]
\centering
\includegraphics[width=1.0\columnwidth]{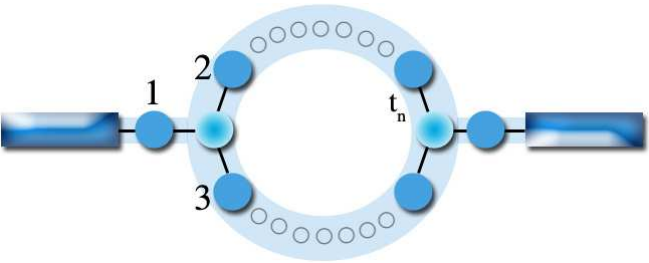}
\caption{
(Color online) The ring attached to two leads can be modeled as two three-way junctions. Each of this three-way junctions connects to one lead and the subsections corresponding to each of the arm in the ring.
}
\label{f:ring2}
\end{figure}
In a ring system, the point where a lead and a ring subsection join can be considered as a three-way junction, as illustrated in Fig. \ref{f:ring2}. In this section we show how this method works in one dimensional structures but it can be easily extended to more dimensions. We first find the boundary condition for a three-way junction by assuming that the wave function used in the scattering S-matrix is also the eigenfunction of the system.
Generally, the lattice distance in the three leads are $a_1$, $a_2$, and $a_3$, with 
$t_n=\hbar^2/2m_na_n$ and $n=1,2,3$. The potential at the joint point, where the three leads come together, is $t_0$. Under this assumption, the eigenfunction and eigenenergy of the electron in each lead can be obtained except at the joint point. Our aim is to find the boundary condition that will combine the three eigenfunctions in the leads and at the joint
point to obtain the wave function for the whole system. Let us assume that the three eigenfunction corresponding to the three leads are $\Psi_1, \Psi_2$ and $\Psi_3$ and the function at the joint point is $\Psi(0)$. Thus, we have four equations:
\begin{eqnarray}\label{boundary1}
(E-2t_n)\Psi_n(a_n)+t_n\Psi_n(2a_n)+t_n\Psi(0)=0
\end{eqnarray} 
\begin{eqnarray}\label{boundary2}
(E-2t_0)\Psi(0)+ \sum_{i=1,2,3} t_i\Psi_i(a_i)=0
\end{eqnarray}
where $n=1,2,3$ represents each of the leads. Since $\Psi_n$ satisfies 
$(E-2t_n)\Psi_n(a_n)-t_n\Psi_n(2a_n)-t_n\Psi_n(0)=0$ and comparing this
with Eq. (\ref{boundary1}), we obtain the first boundary condition:
\begin{eqnarray}\label{bc1}
\Psi_1(0)=\Psi_2(0)=\Psi_3(0),
\end{eqnarray}
which is equivalent to the condition that the wave function is
continuous at the boundary. As a result, Eq. (\ref{boundary2}) can be
rewritten as
\begin{eqnarray}\label{bc2}
&&(E-(2t_0-t_1-t_2-t_3))\Psi(0)\nonumber
 \\&&+t_1(\Psi_1(a_1)-\Psi_1(0))+t_2(\Psi_2(a_2)-\Psi_2(0))\nonumber
\\&&+t_3(\Psi_3(a_3)-\Psi_3(0))=0
\end{eqnarray}
and we see that Eq. (\ref{bc2}) is equivalent with saying that the first derivative is continuous except for a $\delta$ function at the junction point.\cite{Jian-Bai Xia}

Using these boundary conditions in the tight binding model, the S-matrix of a three-way or multi-way junction can be calculated. It should be noted that the matrix we directly calculate is the so called S'-matrix which is not unitary.\cite{datta} The
relation between the S-matrix and the S'-matrix can be given as
\begin{eqnarray}\label{S'-S}
S_{mn}=S'_{mn}\sqrt{\frac{v_m}{v_n}},
\end{eqnarray}
where $v_{m,n}$ is the velocity of the particle in the $m(n)$-th lead. The ring system attached to two leads can be considered as two three-way junctions each of which connects to one lead. These two junction S-matrices can be combined using the boundary conditions (Eqs. \ref{boundary1} and \ref{boundary2}). With both S-matrices, the transmission function can be calculated by combining S-matrices with the advantage that we are able to see the contribution of each Feynman paths, as outlined in Ref. \onlinecite{datta}. 
Having computed this S-matrix under proper boundary conditions, we can use it recursively to compute the transmission probabilities
in order to study the effects of the transparent lead approximation done in prior analytical works.
\section{Effects due to the heavy-hole nature of the carriers}\label{hh_vs_e}
\subsection{Aharonov-Casher Effect}
\begin{figure}[!t]
\begin{center}
\includegraphics[width=0.45\textwidth]{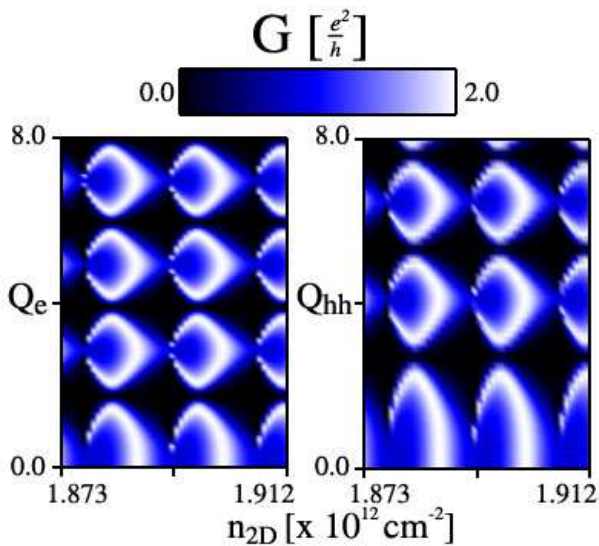}
\caption{
(Color online) Calculation of the zero temperature conductance in a single-moded ring connected to two leads based on the Landauer-B\"{u}ttiker formula. The conductance in both electron and heavy-hole systems is modulated as a function of the spin-orbit coupling and the carrier concentration. In the left panel we present the result for an electron system and in the right panel for a heavy-hole system; both systems have an effective mass of $m^* = 0.031 ~ m_0$.
}
\label{f:1}
\end{center}
\end{figure}
Approximate analytical forms for the conductance in ring structures have been found for electrons\cite{APL_75_695,PRB_69_235310,PRB_69_155335} and heavy-hole systems:\cite{Kovalev:2007_a}
\be \label{e:linear}
G = \frac{e^2}{h} \left[1 - \cos \left(\pi \sqrt{1 + Q_e^2}  \right) \right] 
\ee
\be \label{e:cubic}
G = \frac{e^2}{h} \left[1 - \cos \left(3 \pi \sqrt{1 + \frac{Q_{hh}^2}{9}} \right) \right]  
\ee
 The key approximation of the formulas above is that the coupling between the lead and the ring is perfectly transparent, neglecting backscattering effects that may lead to resonances and other self-interference effects. Another assumption is that the transport of the quasiparticles is just from one lead to the other, traveling in one of the two arms only once, i.e. it does not take into account that the quasiparticles could wind around the ring structure more than half the ring's circumference. Eq. (\ref{e:cubic}) is derived under an the assumption of a thin channel ring ($k_{F} w \lesssim 1$).\cite{Kovalev:2007_a}

In Fig. \ref{f:1} we show the contour plot of the zero temperature conductance in a single moded rings as a function of both the electron density and the dimensionless SO interaction strength using the full Landauer-Buttiker formalism. The left plot  corresponds to electrons traveling in the ring  (uses the Hamiltonian in Eq. (\ref{H_1})). and the right plot corresponds to the  heavy-hole system (uses the Hamiltonian in Eq. (\ref{H_3_simple})). 
 In this and all subsequent calculations we considered a ring of radius $r_0 = 1 \mu m$, an effective mass $m^* = 0.031 ~ m_0$, and starting from zero, we ramp up the dimensionless parameter controlling the strength of the SO splitting energy to 8. In the plots presented in figure \ref{f:1}, we vary the carrier concentration from $n_{2D} = 1.873 \times 10^{12} cm^{-2}$ to $n_{2D} = 1.912 \times 10^{12} cm^{-2}$.  These parameters correspond to the system measured in Ref. \onlinecite{Konig}. Further, we have chosen the carrier concentration assuming an infinite two-dimensional (2D) gas, the situation in the semi-infinite leads which are the particle reservoirs. As is evident in Fig. \ref{f:1}, the character of the SO coupling of the particles traversing the ring is of importance. As a function of the dimensionless parameter $Q_e /Q_{hh}$ the heave holes start at a slower rate reaching the same rate at higher values as the electron system. However, in experiments, where these parameters are a function of the gate voltage, we speculate that the oscillations as a function of gate voltage will  likely  be higher in the thin ring heavy hole system since the effective spin-orbit coupling is enhance with respect to the bulk value due to the confinement. 

A comparison of the relative oscillation frequency in experiments, assuming a splitting proportional to the top gate voltage in most set-ups where spin-orbit coupling can be tuned, would require a translation of the above figures to those parameters together with an accounting of the rate of change of the carrier density with gate voltage. For typical parameters, ignoring the carrier dependence on the oscillations, 
same parameters for electrons give a slower frequency of oscillation in the conductance as compared to the heavy-holes system when ($k_{F} \lesssim 1$).\cite{Kovalev:2007_a}  

Our calculation is in partial agreement with the analytical formulas in that the conductance is modulated by the strength of the SO parameter and also exhibits, at certain values of $Q$ (but independent of the particle density, i.e. Fermi energy), a zero conductance. As Souma and Nikoli{\'c}\cite{Souma} pointed out, these null values correspond to the zeroes of Eq. (\ref{e:linear}) and in the heavy-hole case to zeroes of Eq. (\ref{e:cubic}). Similarly, the numerical results of Moln{\'a}r and coworkers\cite{PRB_69_155335} found a dependence of the conductance in the Fermi wave-vector. These conductance oscillations as a function of the particle concentration not captured by the analytical treatment are due the neglect of backscattering from the leads and related to resonance energies of the rings as we show below.

\begin{figure}[!t]
\begin{center}
\includegraphics[width=0.45\textwidth]{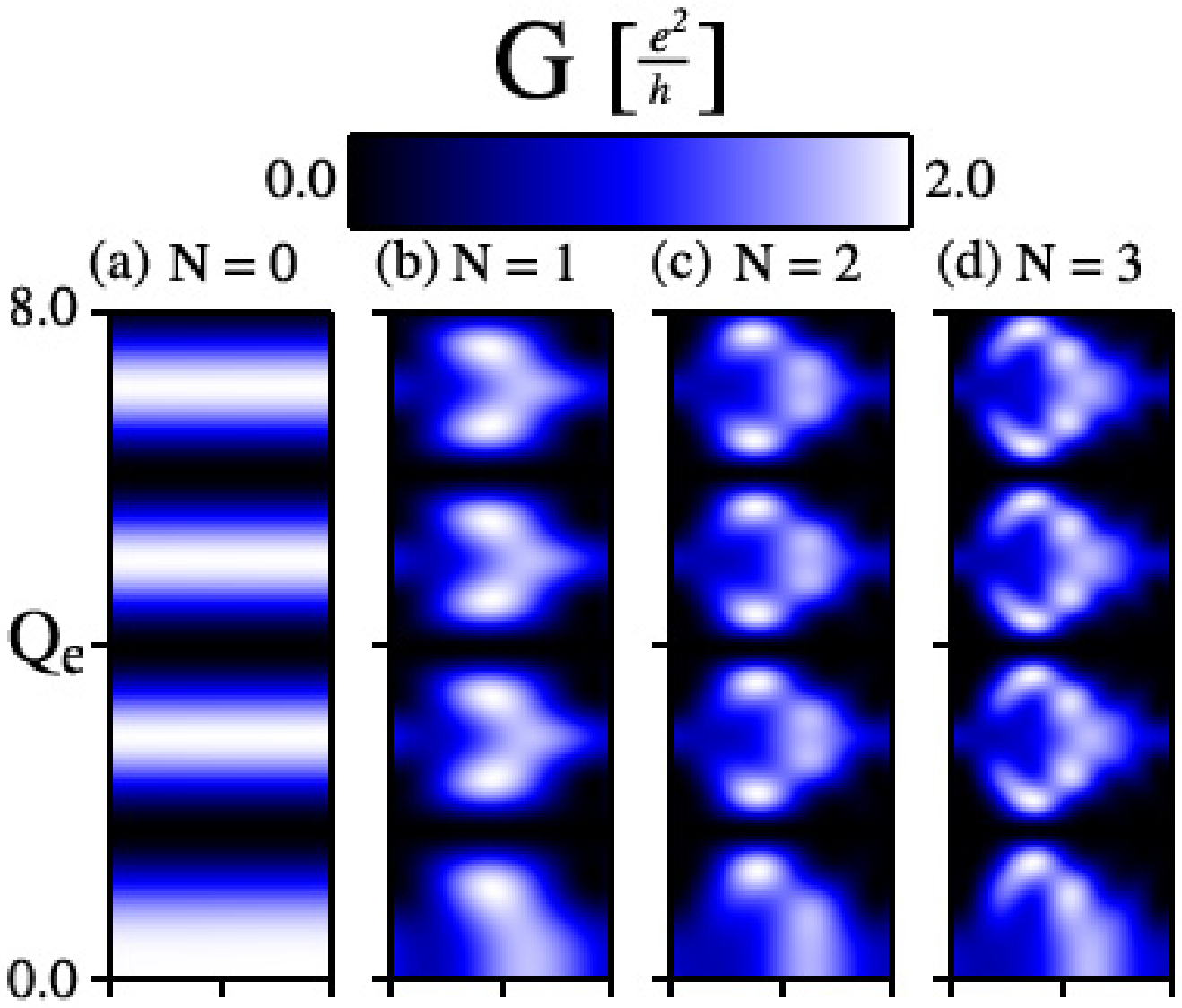}

\includegraphics[width=0.45\textwidth]{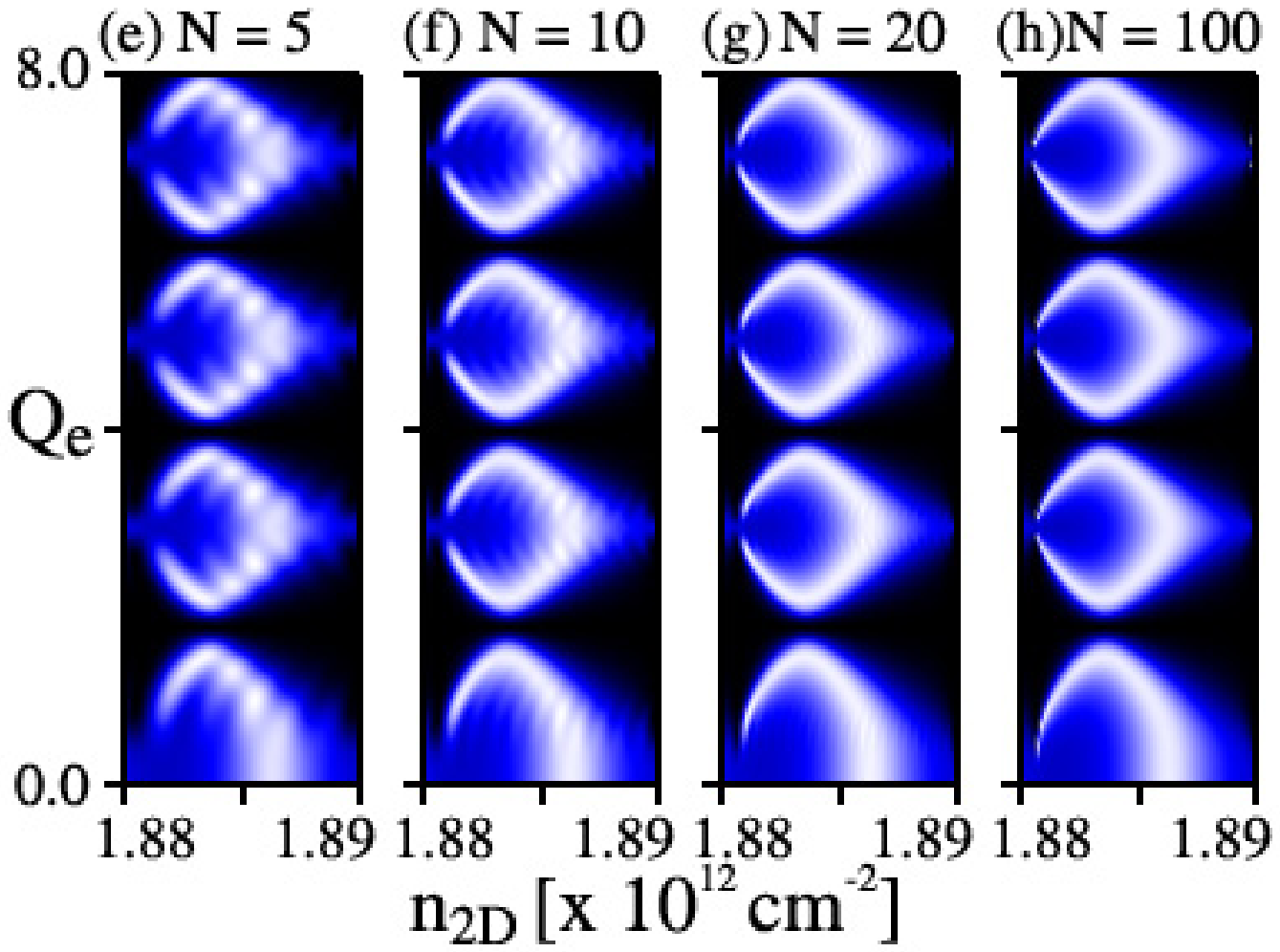}
\caption{
(Color online) Calculation based on the tight-binding model of the zero temperature conductance in a single-moded ring connected to two leads using a recursive S-matrix method. Each panel shows the conductance modulation as a function of both the spin-orbit interaction and the carrier density of the electrons when backscattering from the exit lead $N$ times. $N\rightarrow \infty$ corresponds to the full calculation shown in Fig. \ref{f:1}.
}
\label{f:2}
\end{center}
\end{figure}

Using a recursive S-matrix method we have performed calculations assuming the boundary condition tight-binding model to study the effects of backscattering at the leads and the validity of the transparent leads approximation. With this method we are able to obtain the conductance for different paths taken by the particles. In Fig. \ref{f:2} we present contour plots of the conductance as a function of both the electron density and the dimensionless SO interaction strength for the same electron system as in Fig. \ref{f:1} while increasing the number of allowed backscattering events. In this calculations, one lead is used for injecting electrons and the other will allow the electrons to exit the ring once they completed at most $N$ scattering events from that same lead. The transparent leads case corresponds to Fig. \ref{f:2} (a), where no backscattering is allowed in  exact agreement with Eq. (\ref{e:linear}). If we allow for at most a single scattering event at the right lead before exiting the structure while accounting for all such possible paths, we obtain the contour plot in Fig. \ref{f:2} (b). The continuous structure now gives way to a conductance that depends on the carrier density and it is further divided into periodic sub-structures. Each of these sub-structures shows that as a function of the SO coupling the conductance has either one or two peaks, depending on the carrier density. If the carrier concentration is chosen so that the Fermi energy is in resonance with an eigenenergy of the discretized ring then there is only one peak in the substructure and it happens at the same maximum value as when the lead was transparent. When the carrier density is chosen so that the Fermi energy is not very close to a value of the eigenvalues of the ring then the electron gets backscattered once and as it travels around the ring the path taken and the same path but time-reversed are interfering with each other, similar to weak localization corrections to the conductance.\cite{weak_loc} 
The $Q_e=0$ result is also in agreement with the results obtained by Lucignano {\it et. al.}\cite{Lucignano} for AB rings where once they allowed scattering at the leads, the Fourier spectrum of the magneto-conductance displays  an additional peak at double the fundamental frequency ($2/ \Phi_0$).

The rest of the panels show the result of more backscattering events being allowed. Finally, the periodic structures seen with this "path" calculations get closer to the structures found using the full Landauer-B\"{u}ttiker calculation and these two methods are the same in the limit where the number of allowed backscattering events is taken to infinity. One thing we need to note is that in the heavy-hole calculation, as $Q_hh$ is increased there is a small curvature followed by the sub-structures in the SO coupling axis. The reason for that deviation is a numerical artifact. We have repeated the calculation varying the number of lattice points used to represent the discretized ring and as more points are used the sub-structures shift is reduced.  This curvature is apparent in previous numerical calculations\cite{Souma} but not recognized as numerical artifact. In our calculations we have chosen the carrier concentration such that the Fermi energy of the system is close to the bottom of the band and hence model in this way the continuous effective mass model appropriate for these semiconductor systems. 
%
\subsection{Spin Hall Effect}

\begin{figure}[!t]
\begin{center}
\includegraphics[width=0.45\textwidth]{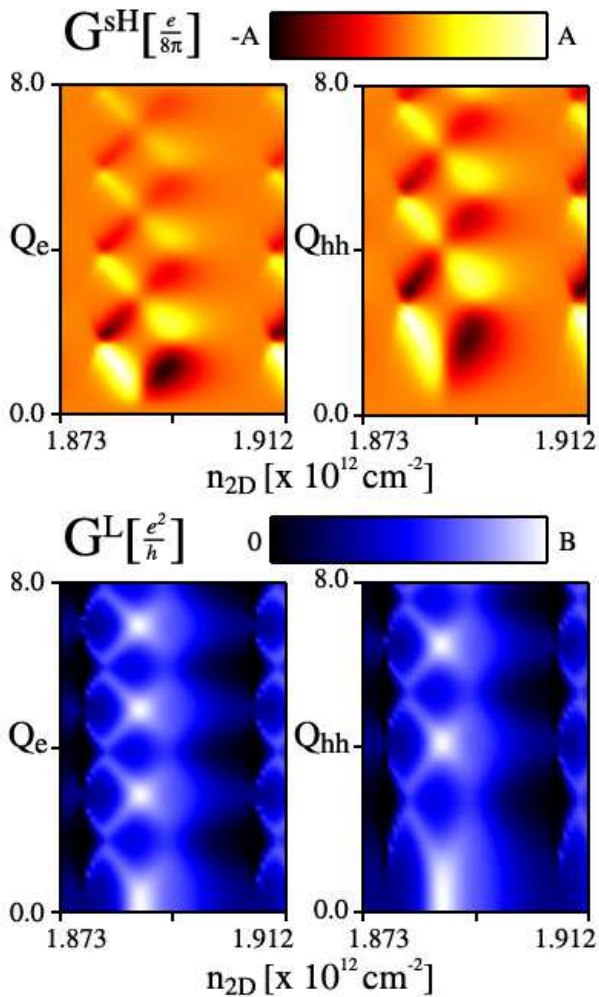}
\caption{
(Color online) Calculation of the zero temperature spin-Hall and longitudinal conductance in four terminal ring structure based on the Landauer-B\"{u}ttiker formula. The spin-Hall conductance (top panels) and longitudinal conductance (bottom panels) in electron and heavy-hole systems is modulated as a function of the spin-orbit coupling and carrier concentration. We present on the left the electron system where the spin Hall conductivity has a maximum value of $A = 0.72 \times e/(8 \pi)$ and the longitudinal conductance has  a maximum value of $B = 1.2 \times e^2/h$. In the heavy-hole system, shown on the right, the maximum value of the spin Hall conductivity is $A = 1.0 \times e/ (8 \pi)$ and the longitudinal conductance reaches a maximum value of $B = 1.7 \times e^2/h$.
}
\label{f:sHe_e_hh}
\end{center}
\end{figure}

The proposal of intrinsic spin Hall effect\cite{SHE1} and the experimental observation of the effect (both extrinsic and intrinsic) \cite{SHE2} has generated a lot of interest in the semiconductor spintronics community. Numerical calculations based on the Landauer-B\"{u}ttiker formalism on quantum coherent ballistic rings have shown that the spin Hall conductance exhibits quasi-periodic oscillations as a function of the Rashba SO coupling.\cite{Souma:2004_a} The calculation found several interesting predictions including the possibility of generating a spin Hall current in a ring with two longitudinal probes that inject/extract current and two transverse probes that would measure the voltage. Although experimental realization of such four-probe rings could prove difficult, we are interested in how the cubic SO term would affect the spin Hall effect. 

As we show in Fig. \ref{f:sHe_e_hh}, there are oscillations in the spin Hall conductance as the carrier density is increased. This oscillations have similar nature as those found in the conductance of the two-probe rings. The pattern seen by Souma and Nikoli{\'c}\cite{Souma:2004_a} is recognized from the figure, as we also observe that the spin Hall conductance oscillates in sign and that as the SO interaction increases the oscillations in the spin Hall conductance dampen. The longitudinal conductance of the four-probe ring shows common features with the conductance of the two-probe ring with two distinct features: (1) it does not vanish at specific values of the SO interaction and (2) the maximum value of the conductance is now lower than $2e^2/h$ due to the dephasing effects of the additional voltage probes. The addition of the transverse leads has shifted the structures to the left and now the structures we see are overlapping. It is at this overlap that the spin Hall conductance shows the x-like structures.  The parameters used in these plots are similar to those used in the two-probe rings: the effective mass $m^* = 0.031 ~ m_0$ is the same in both systems, we vary the carrier concentration from $n_{2D} = 1.873 \times 10^{12} cm^{-2}$ to $n_{2D} = 1.912 \times 10^{12} cm^{-2}$, and starting from zero, we ramp up the dimensionless parameter controlling the strength of the SO splitting energy  to 8.
 We note that the strength of the signal is larger for the heavy-hole system: maximum spin Hall conductance is $G^{sH}_e = 0.72 \times e/(8 \pi)$ for electrons and $G^{sH}_{hh} = e/ (8 \pi)$ for heavy-hole system and the longitudinal conductance has a maximum value of $G^{L}_e = 1.2 \times e^2/h$ and $G^{L}_{hh} = 1.7 \times e^2/h$.  As explained in the previous section, the curvature followed by the sub-structures as SO increases in the heavy-hole plots is a numerical artifact.

\section{Inhomogeneous Spin-Orbit coupling}\label{inhomo}
\begin{figure}[!t]
\begin{center}
\includegraphics[width=0.45\textwidth]{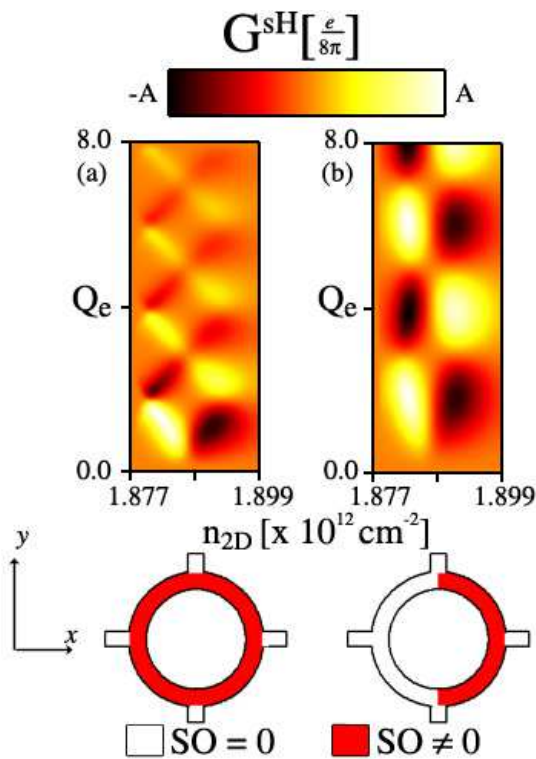}
\caption{
(Color online) Calculation of the zero temperature spin-Hall conductance (four terminal device) based on the Landauer-B\"{u}ttiker formula. The spin-Hall conductance in electron systems is modulated as a function of the spin-orbit coupling and electron density. When the spin-orbit coupling is homogeneous (Left picture), the spin Hall conductivity has a maximum value of $A = 0.72 \times e/(8 \pi)$. When the spin-orbit coupling is present in only one half of the ring structure (Right picture), the maximum value of the spin Hall conductivity is $A = 1.07 \times e/ (8 \pi)$.
}
\label{f:sHe_full_half}
\end{center}
\end{figure}

Experimental efforts in mesoscopic rings have manipulated the SO splitting (in both the ring structure and the leads) via a gate voltage. Since the SO interaction depends on the surface electric field, there is a natural interest in studying the effects of a gate that covers only part of the system.\cite{PRL_92_086602,PRB_76_085319} 
Tserkovnyak and Brataas have predicted and enhancement of the interference effects in mesoscopic rings with inhomogeneous SO splitting for weak SO coupling systems.\cite{PRB_76_155326} We study those effects with our numerical techniques in spatially varying SO coupling systems. We observe an enhancement in the spin Hall conductivity in the four-probe rings and an unexpected modulation of the conductivity in the two-probe rings. 

\begin{figure}[!t]
\begin{center}
\includegraphics[width=0.45\textwidth]{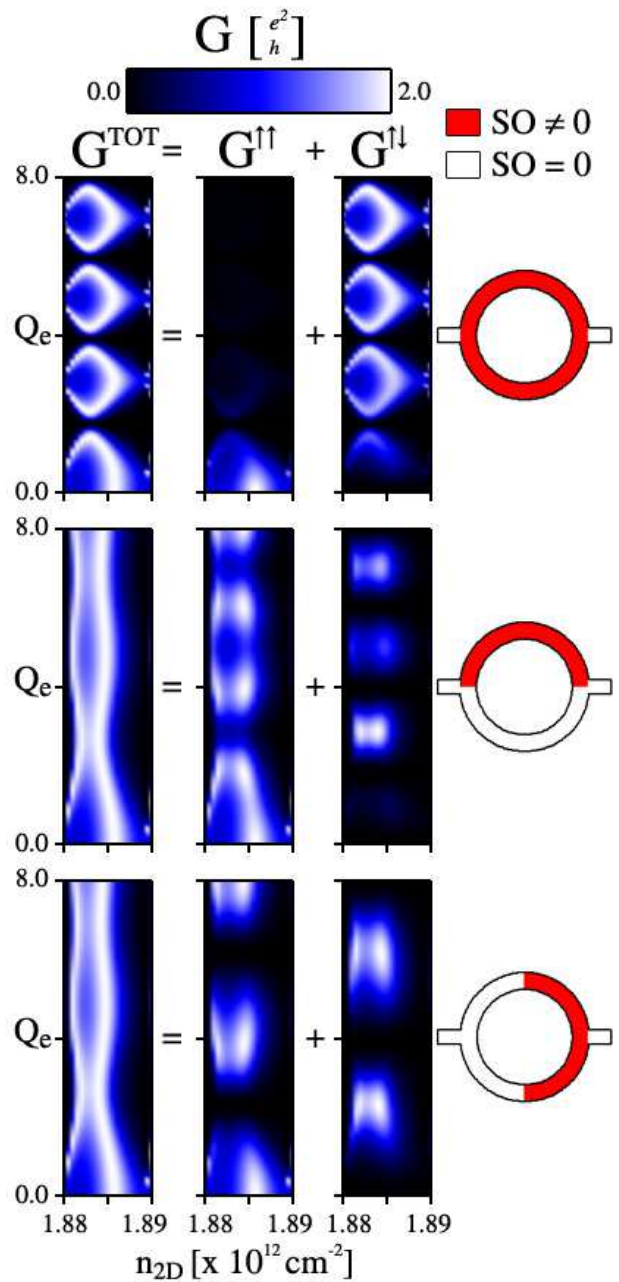}
\caption{
(Color online) Calculation of the zero temperature spin-resolved conductance based on the Landauer-B\"{u}ttiker formula.
}
\label{f:G_full_half}
\end{center}
\end{figure}

In the last two figures, we present the conductivities for a ring with a constant SO coupling ($\alpha_o$) and compare it to the conductivities obtained in a ring where the electrostatic gate covers half of the ring. The SO coupling depends on the position as: 
\bea
\alpha(x) &=& \frac{\alpha_o}{2}\left(1 + \tanh \frac{x - x_o}{\Delta} \right) 
\eea 
Although not shown in the sketches in Fig. \ref{f:sHe_full_half} and \ref{f:G_full_half}, the value of $\Delta$ in the calculation allows for a transition region in the range of one tenth of the radius of the ring. This transition region would account for the electric field from the gate affecting parts of the ring not covered by the gate. Previous studies have shown that if the SO interaction is turned on abruptly there are strong scattering effects in the two regions.\cite{Nikolic:2004_a,PRB_73_115342} We have found that the sharpness of the transition region does not significantly change the conductivity patterns apart from the noticeable effect of a higher 'effective' SO coupling along the ring. 

Fig. \ref{f:sHe_full_half} shows the spin Hall conductivity as a function of $Q_e$ and carrier density for a fully covered and a partially covered configuration. The oscillations are slower in the partially covered one due to the fact that the configuration of the right ring has a lower effective SO coupling. Yet, the pattern is significantly stronger ($\sim 35 \%$). Also important is that the spin Hall conductivity does not dampen as the SO interaction is raised. 

Next we study the spin-dependent conductivity for inhomogeneous SO interaction. The configuration we consider for the measurement would require the injection of spin polarized electrons into the ring structure and a means to detect their polarization as they exit the ring. Both the spin injection and the detection of the spin direction could be probed by electrical\cite{Nat.P_3_197} or optical\cite{Nat.P_3_542} means with current experimental techniques. 

In Fig. \ref{f:G_full_half}, we show the contour plot of  zero temperature conductances (total and spin resolved) in single moded rings as a function of both the electron density and the dimensionless SO interaction strength for three different spatial configurations of the SO interaction. In the top panel we show the homogeneous case, which has been previously analyzed.\cite{Souma} As the SO increases from zero, the injected electrons flip their spin direction. While the total conductance has the familiar periodic structure, these oscillations are only present in the spin-flipped (spin up injected on left lead and spin down detected on the right lead) conductivity. The spin conductivity for spin-conserved components (spin up injected and spin up detected) decays rapidly. The next two panels present the same calculation when only half of the ring is covered by the gate. As we see, the total conductivity is devoid of null values and is identical for the two configurations. In contrast, the spin-flipped and the same-spin conductivity show a marked difference from the homogeneous SO case. The spin-conserved conductivity does not decay as the SO increases but rather oscillates. Meanwhile, the spin-flipped conductivity is also oscillating and both conductivities contribute to the total value equally. This behavior illustrates that the reversal of the spin polarization could be controlled in ring structures by changing the voltage in a top-gate.

\section{Summary}\label{conc}
We have studied the quantum interference effects induced by the Aharonov-Casher phase in asymmetrically confined two-dimensional electron and heavy-hole ring structure systems taking into account the electrically tunable SO interaction. We calculated the non-adiabatic transport properties of charges in ring structures and confirm the analytic result.\cite{Kovalev:2007_a} 
We have found that the interference effects depend both on the SO splitting of the bands and the carrier density. 
Further, we are able to show that the contribution to the conductivity of non-transparent leads affect the conductivity and fully explains the  Landauer-B{\"u}ttiker result dependence on carrier density.
We have calculated the spin Hall conductivity and longitudinal conductivity in four-probe rings. Our analysis suggests that for hole doped systems both conductivities are stronger than the conductivities found in electron doped system. Finally, we have investigated the conductance of mesoscopic rings with spatially inhomogeneous SO coupling. In this case the spin Hall conductivity oscillates in a similar fashion as in the homogeneous SO case but, as the gate voltage is increased, the signal strength does not weaken in contrast to the homogeneous case. We have also found that devices with inhomogeneous SO interaction exhibit intriguing spin resolved conductivities which could lead to the modulation of the spin direction of polarized carriers. 
\section*{Acknowledgment}
The authors thank Y. Tserkovnyak and L.P. Z\^{a}rbo for stimulating discussions.
This work was supported
support from
ONR under Grant No. ONR-N000140610122, by NSF
under Grant No. DMR-0547875, by SWAN-NRI, by
EU No. IST-015728, Czech Republic No. AV0Z1-010-914,
No. KAN400100625, No. LC510, and No. FON/06/E002.  Jairo Sinova is a
Cottrell Scholar of the Research Corporation.

\end{document}